\documentclass[manuscript,screen]{acmart}
\acmConference{}{}{}

\makeatletter %1
\def\mdseries@tt{m} %1
\makeatother %1
\usepackage[plain]{fancyref}
\usepackage[draft=true]{minted} %2
\usepackage{color}
\usepackage{hyperref} %5
\hypersetup{
    colorlinks=true,
    linkcolor=blue,
    filecolor=red,      
    urlcolor=magenta,
    breaklinks=true,            %3
}
\usepackage{booktabs} % For formal tables
\usepackage[edges]{forest}

\definecolor{folderbg}{RGB}{124,166,198}
\definecolor{folderborder}{RGB}{110,144,169}

\def\Size{4pt}
\tikzset{
  folder/.pic={
    \filldraw[draw=folderborder,top color=folderbg!50,bottom color=folderbg]
      (-1.05*\Size,0.2\Size+5pt) rectangle ++(.75*\Size,-0.2\Size-5pt);  
    \filldraw[draw=folderborder,top color=folderbg!50,bottom color=folderbg]
      (-1.15*\Size,-\Size) rectangle (1.15*\Size,\Size);
  }
}

\forestset{is file/.style={edge path'/.expanded={%
        ([xshift=\forestregister{folder indent}]!u.parent anchor) |- (.child anchor)},
        inner sep=1pt},
    this folder size/.style={edge path'/.expanded={%
        ([xshift=\forestregister{folder indent}]!u.parent anchor) |- (.child anchor) pic[solid]{folder=#1}}, inner xsep=0.6*#1},
    folder tree indent/.style={before computing xy={l=#1}},
    folder icons/.style={folder, this folder size=#1, folder tree indent=3*#1},
    folder icons/.default={12pt},
}

\AtBeginDocument{%
  \providecommand\BibTeX{{%
    \normalfont B\kern-0.5em{\scshape i\kern-0.25em b}\kern-0.8em\TeX}}}

\setcopyright{none} % MICHELE: remove this

\acmDOI{xx.xxx/xxx_x}

\begin{document}
\sloppy % MICHELE: remove
\title{Towards a secure API client generator for IoT devices}
% MICHELE \title{cpp-tiny-client: A secure API client generator for IoT devices}

\author{Anders Aaen Springborg}
\email{asprin18@student.aau.dk}
\affiliation{%
  \institution{Aalborg University}
  \city{Aalborg}
  \country{Denmark}
}
\author{Martin Kaldahl Andersen}
\email{martan18@student.aau.dk}
\affiliation{%
  \institution{Aalborg University}
  \city{Aalborg}
  \country{Denmark}
}
\author{Kaare Holland Hattel}
\email{khatte18@student.aau.dk}
\affiliation{%
  \institution{Aalborg University}
  \city{Aalborg}
  \country{Denmark}
}
\author{Michele Albano}
\email{mialb@cs.aau.dk}
\affiliation{%
  \institution{Aalborg University}
  \city{Aalborg}
  \country{Denmark}
}

% MICHELE \renewcommand{\shortauthors}{Springborg A.A., et al.}

\begin{abstract}
Given the success of IoT platforms, more developers and companies want to include the technology in their portfolio. However, in the case of single board microcontrollers, the support for networking operations is not ideal, and different IoT platforms allow access to the networking submodule via different libraries and system calls, leading to a steeper learning curve. 
Code generators for API clients can enhance productivity, but they tend to generate universal purpose code, and on the other hand the networking primitives of IoT devices are platform specific, especially when security mechanisms such as Transport Layer Security are part of the picture. 
This paper presents \texttt{cpp-tiny-client}, an API client generator developed as a plugin for the OpenAPI Generator project, which can tailor the generated code based on the IoT platform specified by the user. Our work allows to generate correct code for API clients for IoT devices, and thus can empower a developer with more productivity and a faster time-to-market for its own applications. 
By combining together mainstream technologies only, \texttt{cpp-tiny-client} offers a gentle learning curve. Moreover, experiments show that the generated code has a reasonable footprint, at least with respect to the IoT devices that were used in the validation of the work. 
The code related to this work is available through the OpenAPI Generator project~\cite{OpenAPIGenerator}. 
This technical report is an extension of~\cite{acmsac22}, and it integrates the information presented at the ACM SAC 2022 conference. 
\end{abstract}




\maketitle

\section{Introduction}

Year by year Internet of Things (IoT) applications are becoming increasingly popular, and all the market segments related to IoT show no sign of slowing down~\cite{columbus2017roundup}. 
Common use cases span from monitoring industrial machines~\cite{albano2019mantis} to hobbyist projects and home automation~\cite{javed2016building}. 
However, industry acceptance is not as fast as it could be, and {\em "Security, IT/OT integration and unclear ROI are the greatest barriers to IoT adoption today."}~\cite{columbus2018roundup}

One fundamental feature of IoT devices is the capability of communicating over the Internet, and in fact most IoT definitions~\cite{berte2018defining} are focused on this characteristic. 
The plethora of use cases of the IoT raises the need for both an easy way of sending and receiving data across the internet, but also for doing it securely, since the steeper barrier for IoT market acceptance is in fact security~\cite{columbus2018roundup}. 
Luckily, most modern IoT devices come with built-in internet connectivity. Augmented with a proper support by the operating system, this can take care of the lowest layers of the ISO/OSI stack~\cite{palattella2012standardized}. 

Even though current research trends are looking into novel security techniques (see for example~\cite{frimpong2020secon}), Transport Layer Security (TLS) is the de facto standard technology for keeping an
Internet connection secure and safeguarding any sensitive data that is being sent between two systems~\cite{SSL_TLS}. The protocol requires two cryptographic keys, one to be shared publicly and one to be kept private. A client needs to know in advance the public key for the specific server in order to verify its identity and set up a secure connection, or as an alternative a public key certificates can be sent from the server to the client. In the latter case, which is the most common, the client can verify that the certificate is valid according to a Certification Authority (CA) that installed a root certificate on the client beforehand, thus validating the identity of the server, and then the client can extract the public key of the server from the issued certificate to start communicating with it securely.~\cite{tlsmastery}

With regards to the application layer, many modern systems use Representational State Transfer (ReST) Application Programming Interfaces (APIs) for the communication between server and client, since it provides many benefits such as interoperability~\cite{costa2014evaluating}. Most often than not, a REST API is built upon the Hypertext Transfer Protocol (HTTP)\cite{yannakopoulos2003hypertext}, and on its secure sibling HTTP Secure (HTTPS), which exchanges HTTP messages and protects them using TLS before sending it over the internet. 
When an IoT device is part of the scenario, support for HTTPS must take into account that servers use different certificate providers, and the client needs to store the corresponding root certificate to verify the identity of the server and build up a secure connection. This can represents a problem since there is no guarantee that there is not enough flash memory for all the existing root certificates in the IoT device's flash memory. 

Software developers or companies who implement API clients on IoT devices often develop the code for a specific microcontroller, since the HTTP/HTTPS libraries are microcontrollers-specific and for example the code developed for a ESP32~\cite{Esp32HTTPClient} cannot be used "as it is" on a ESP8266~\cite{Esp8266HTTPClient}. 
Code generation can improve developers' productivity and reduce time-to-market of applications, thus it can be an attractive option for software developers and companies alike. 
However, existing code generators for microcontrollers do not support networking very well. In fact,
these generators focus on the generation of more universal purpose code, but on the other hand the networking primitives of IoT platforms are hardware specific.

%% Goal
%This paper presents 
This report describes the creation of \texttt{cpp-tiny-client}, an API client generator tailored for IoT devices, which was developed as a plugin for the OpenAPI Generator project\cite{OpenAPIGenerator}. The code generator allows to establish both HTTP and HTTPS connections with the server, and it supports potentially multiple CA root certificates -- if the microcontroller's flash memory is large enough. 

The report is structured as follows. Section~\ref{sec:background} provides background information on microcontrollers, code generation and related tools, and culminates by selecting which of them are taken into account into the development work -- and why. Section~\ref{sec:design} and Section~\ref{sec:implementation} describe the design and the implementation of the \texttt{cpp-tiny-client} generator respectively, the first being useful for a user of \texttt{cpp-tiny-client}, and the second for a developer interested into extending it to more microcontrollers. Section~\ref{sec:validation} documents the validation process used for the \texttt{cpp-tiny-client}, with a particular focus on the memory footprint. Section~\ref{sec:limitation} concludes the paper and proposes future work to overcome current limitations.

%%%%%%%%%%%%%%%%%%%%%%%%%%%%%%%%%%%%%%%%%%%%%%%%%%%%%%%%%%%%%%%%%%%%%%%%%%

\section{Background Information}\label{sec:background}

This section describes building blocks that could be put together to create a code generator solution for IoT devices. 
Later on, the section surveys existing approaches to code generation, not only to compare with our proposed idea, but also to identify which one to extend to realize \texttt{cpp-tiny-client}. 
Finally, this section concludes by taking decisions on what to integrate to design and implement \texttt{cpp-tiny-client}.

\subsection{Which IoT platform and tools?}\label{app:platform}

The IoT platform and its development environment have an impact on how the (generated) code must be structured and will be deployed to the IoT devices. 
Even though many microcontrollers hit the market every year, the IoT platforms that are leading the market are not so many. 
Arguably, the Arduino platform is very popular, and well supported by many IoT devices manufacturers. For example, Espressif Systems creates popular microcontrollers~\cite{EspressifWorldLeading} that support two different frameworks, Arduino and ESP-idf. 
The following discussion revolves around the Arduino Integrated Development Environment (IDE) and PlatformIO, which are good development environment candidates for \texttt{cpp-tiny-client}.

\subsubsection{Arduino IDE}
\cite{ArduinoIDE}
Developing a project for Arduino is often done with the Arduino IDE. It is open source and requires little setup. To create a project, the user selects its Arduino model, which serial port it is connected to, and can start coding. While this process is simple, it also lacks some essential features for developing a larger project in a group of people. Primarily, Arduino IDE has no version control integration, no code debugger, and no intellisense. Arduino IDE projects are developed in the Arduino language, which is a subset of C++. 

\subsubsection{PlatformIO}
\cite{PlatformIO} is a {\em "cross-platform, cross-architecture, multiple framework, professional tool for embedded systems engineers and for software developers who write applications for embedded products"}~\cite{WhatIsPlatformIO}. PlatformIO supports many different platforms and frameworks, including Arduino. This tool allows the option of being integrated in various IDEs, however, on its website it is recommended to use it via the PlatformIO Visual Studio Code extension. 
A PlatformIO development project includes a \texttt{platformio.ini} file where the environment, library dependencies, etc.\ are specified. PlatformIO projects are developed in the C++ language.

\subsection{Which IoT microcontrollers?}\label{app:mc_compare}
Espressif Systems creates popular microcontrollers \cite{EspressifWorldLeading} that support two different IoT platforms, one of them being Arduino. 

While many microcontrollers come with very limited amounts of flash memory and RAM, the comparatively larger memory of the ESP family presents the user with more options. 
Table \ref{tab:mcComparison} shows a comparison of the ESP family versus two popular  microcontrollers by Arduino~\cite{ArduinoMegaSpecs, ArduinoUnoSpecs}. Both the ESP32 and ESP8266 come with a much bigger memory pool. This is also the case, when comparing with other popular microcontrollers of the same price range from other vendors. The Arduino Mega also comes without integrated WiFi, but with the option of adding this functionality externally.

\begin{table}[!b]
\resizebox{0.8\linewidth}{!}{%
\begin{tabular}{|c|c|c|c|c|}
\hline
 &
  \textbf{ESP32} &
  \textbf{ESP8266} &
  \textbf{\begin{tabular}[c]{@{}c@{}}Arduino Uno \\ WiFi Rev 2\end{tabular}} &
  \textbf{\begin{tabular}[c]{@{}c@{}}Arduino Mega \\ 2560 Rev 3\end{tabular}} \\ \hline
\textbf{WiFi}                                                   & Yes    & Yes    & Yes    & No    \\ \hline
\textbf{\begin{tabular}[c]{@{}c@{}}Flash\\ memory\end{tabular}} & 1-16MB & 1-16MB & 48KB   & 256KB \\ \hline
\textbf{RAM}                                                    & 512KB  & 160KB  & ~6.2KB & 8KB   \\ \hline
\end{tabular}%
}
\caption{Comparison of popular microcontrollers}
\label{tab:mcComparison}
\end{table}

%%%%%%%%%%%%%%%%%%%%%%%%%%%%%%%%%%%%%%%%%%%%%%%%%%%%%%%%%%%%%%%%%%%%%%%%%
\subsection{Which External Libraries can be Useful?}
Two external libraries appear to cover interesting functionalities to support code generation for API clients. Their purpose is to communicate with a server and send or receive data in JSON format. 

\subsubsection{HTTPClient and ESP8266HTTPClient}
Both libraries are used for making HTTP requests. They have the same interface, but are developed for the ESP32 \cite{Esp32HTTPClient}, and ESP8266 \cite{Esp8266HTTPClient} microcontroller respectively. The libraries are written in C++ and developed by Espressif Systems. They are part of Espressif's Arduino library package, which is based on the Arduino Core library. The libraries are able to establish TLS connections, but they expose a different interface for this use case. HTTPClient needs a root certificate to establish an TLS connection, whereas the ESP8266HTTPClient needs a x509 fingerprint. 

\subsubsection{Steinwurf Bourne}
Bourne is a C++ JSON read/write library developed by Steinwurf~\cite{SteinwurfBourne}. Data is encoded as JSON as it is transferred between  client and  server, and an API client can use Bourne to either convert data that is being sent to the server into a JSON format or parse JSON received from the server into appropriate variables and data structures.

\subsection{Generating Code for IoT Platforms}

%Generating code for IoT devices is not an easy task, especially when the device under consideration is a microcontroller. 
In general, existing API code generators are found wanting with regards to their support for networking for IoT devices, as networking primitives of microcontrollers are hardware specific. 
However, some code generators are easier than others to be integrated in a development process for IoT devices.

\subsubsection{ThingML}
\cite{ThingMLpaper} is a modelling language, with tools for multi-platform code generation~\cite{WhatIsThingML}. The approach involves modeling a distributed system in the ThingML programming language, and using this model to generate code and UML diagrams. ThingML has plugins to communicate across different parts of the distributed system, but no plugin for communicating through HTTP Requests. On the other hand, an example of how to integrate HTTP requests on Arduino in ThingML is shown on their public Github repository \cite{ThingMLServerExample}. It appears that the user has to implement plain Arduino code for networking, and ThingML is wrapped around the plain C code developed manually. This approach does not satisfy the idea of using code generation to simplify the development of code on IoT devices since it requires to have deep knowledge of the particular IoT platform being used and of its networking libraries.

Still, it could be possible to spend significant effort on ThingML to overhaul its design, implement a plugin for network logic, and end up with a reasonable solution for API code generation. However, according to the paper describing ThingML~\cite{WhatIsThingML} and its website, the last drawback of ThingML is that it has only been used by students, and it is not widely used in the industry. 

%provides users with some of the same concepts as the openapi-generator, such as a modelling language and code generators for specific languages \cite{WhatIsThingML}. In addition to the openapi-generator, ThingML supports code generation for a broad range of different domains, besides REST systems. Since the modelling language of ThingML does not explicitly target the domain of REST systems, the modelling language can be used to describe behaviors of systems unrelated to REST, and generate code according to this behavior. The fact that ThingMLs modelling language supports a more general code generation domain, gives the users opportunities to use ThingML for code generation for many different contexts. Where the OpenAPI specification can be described in either YAML or JSON, ThingML has made its own programming language. A ThingML code generator expects a ThingML model, defined in ThingMLs programming language. The fact that a ThingML model must be defined in the ThingML programming language, may be a disadvantage for the users, in terms of simplifying the task of generating API clients for microcontrollers. To use ThingML users must familiarize themselves with this programming language. According to their documentation they " conduct experiments with HTTP used to
%communicate in a REST style"
%. Looking at the HTTP example for arduino in ThingML's public github repository

\subsubsection{CAPSml}
\cite{CASPml} is a model driven code generation framework. This code generator is based on \texttt{CAPS}, which is described as {\em "an architecture-driven modeling framework for the development of IoT Systems"}~\cite{muccini2017caps}. CAPS uses a graphical user interface for the design and modelling. The CAPSml generator uses the software model output from the CAPS software to generate a ThingML language specification of the modeled system. This allows users to generate their systems with the ThingML framework using the graphical design approach of CAPS. CAPSml looks promising since it is designed for IoT development. However, it is based on ThingML, thus it raises its same concerns.

\subsubsection{ArIA}
\cite{ArIA} is another model driven code generator built on top of CAPS. ArIA generates plain Arduino code from a model in CAPS. This allows the user to design behavior in CAPS, which has a graphical user interface, instead of coding directly in the Arduino environment. The ArIA paper shows an example, with a server component, which sends data to a mobile application (illustrated in their paper by Figure 2 of~\cite{ArIA}). The paper does not specify how communication is done on Arduino, and in particular whether it is via a serial port or networking, and if it does this securely. Finally, this code generator inherits the same issues of ThingML, and it lacks documentation. 

\subsubsection{ASM2C++}
\cite{asm2c++} is a tool which can automatically generate C++ code for Arduino. The code is generated based on an Abstract Syntax Machine, more specifically a specification written in Unified Syntax for Abstract State Machine (UASM). The purpose of generating C++ code based on this specification is the rapid design of embedded devices, while retaining a high level of correctness through validation and verification of the syntax. There is no future work planned for generating code for network protocols and sockets, and the URL for downloading the source code of ASM2C++ references to a non existing page. 

\subsubsection{OpenAPI Generator}
\cite{WhatIsOpenAPI}~is an open source project that has the goal of generating server stubs, API clients and API documentation based on an OpenAPI Specification (OAS)~\cite{OpenAPISpecificationDesc}. An OAS allows the user to describe their entire API, including endpoints, input/output parameters for operations, authentication, and more. An OAS can be written in a YAML or JSON format. The OpenAPI Generator project contains API generators for over 40 programming languages, and by the way it supports API client generation for languages commonly used in microcontrollers such as C and C++. 

Users can define an interface with an OAS, and then generate both API clients and server stubs. These components provide corresponding functionality, services, and models as specified, and the generated code can be imported into the users project. The consortium managing and developing OpenAPI Generator is very active, and the documentation of the code generator is great, but the OpenAPI Generator can not generate clients for microcontrollers, since generated code refers to libraries that are not supported by microcontrollers. According to the OpenAPI Generator website~\cite{OpenAPIGeneratorUsers}, the OpenAPI Generator is used by more than 100 companies in the industry, among them Arduino Inc.

\subsubsection{APIGEND}
\cite{laso2020deployment} is built on OpenAPI Generator and allows to generate client APIs for Android and the ESP32 microcontroller. The solution allows the ESP32 to make requests mediated by the Message Queue Telemetry Transport (MQTT) protocol to a server. Albeit interesting, the paper does not mention how security is implemented nor how to adapt the code generator to other IoT platforms, and in particular what to do if the target platform does not have a MQTT library available, or when a MQTT broker cannot be part of the architecture.

\subsubsection{AutoIoT}
\cite{nepomuceno2020autoiot} applies a simplified version of Model Driven Engineering to allow developers without experience in the technique to generate the server-side code to support IoT-based applications. The IoT devices can communicate with the server using MQTT, HTTP or WebSockets, but the solution is directed to the developers of servers that receive data from IoT devices, and it considers that the IoT devices are programmed in the traditional manual way.

\subsection{What to Use for \texttt{cpp-tiny-client}}
The ESP family of microcontrollers was chosen as the first IoT devices to be supported by \texttt{cpp-tiny-client}, because of their popularity, ease of use and connectivity possibilities. For example, they are already used for industrial monitoring i.e.\ in products by \texttt{norvi}~\cite{norviCorp}.
Moreover, while many microcontrollers come with extremely limited amounts of flash memory and RAM, the comparatively larger memory of the ESP family (see Table~\ref{tab:mcComparison}) allows the user to build more complex applications. 
In particular, the focus is on the ESP32 and the ESP8266 microcontrollers, which come with integrated WiFi and allow to perform HTTP and HTTPS requests by means of the HTTPClient~\cite{Esp32HTTPClient} and ESP8266HTTPClient~\cite{Esp8266HTTPClient} libraries respectively. 

For serialization and deserialization of JSON, the Steinwurf's Bourne library was chosen due to its ease of use and because it can be included in a development project as a single C++ file. 
\texttt{cpp-tiny-client} was built on top of the OpenAPI Generator project, since the project is widely used in the industry~\cite{OpenAPIGeneratorUsers}. 
Generated code is compiled using PlatformIO, since it appears more flexible than other development environments for the Arduino microcontrollers.

\begin{figure*}[t!]
    \centering
    \includegraphics[height=6cm]{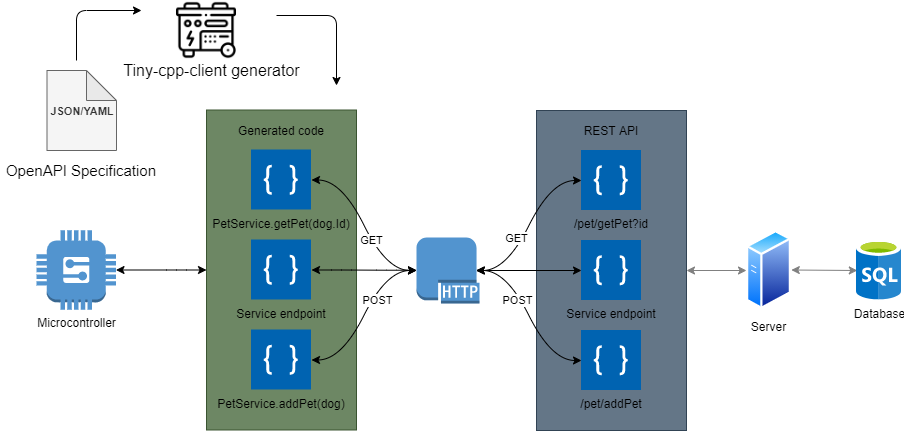}
    \caption{Architecture of client (green) and server (grey) system using the HTTP protocol}
    \label{fig:architecture}
\end{figure*}

\section{Design}\label{sec:design}

%To create an API client, it requires an implementation of both a library which handles the HTTP protocol, and a JSON serialization library. To make the implementation secure, it also requires figuring out how to implement SSL and handling the certificates on devices with a very limited amount of memory. 
\texttt{cpp-tiny-client} was designed to be a plugin for the OpenAPI Generator project, and 
Figure~\ref{fig:architecture} depicts the grand vision for the solution. A microcontroller (on the left) communicates via the HTTP (or HTTPS) protocol with a server, which has access to a Database, e.g. to save received data on permanent storage. 
The green box is the code that is executed on the IoT device, and it is (partly) generated by \texttt{cpp-tiny-client} from the OAS. The grey box is the code executed on the Server, it can also be generated using OpenAPI Generator, but it is not in the scope of this work.

\begin{figure}[t!]
 \centering
    \includegraphics[width=0.5\linewidth]{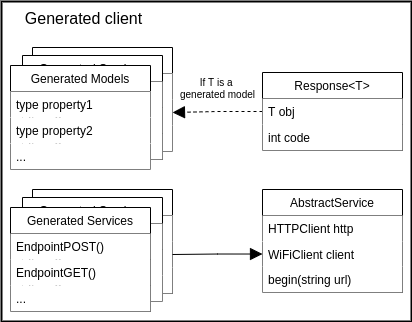}
    \caption{Architecture of generated code}
    \label{fig:abstract_architecture}
\end{figure}

Figure~\ref{fig:abstract_architecture} shows an abstract representation of the code of the generated API client. In OpenAPI Generator's terms, the {\em Generated Models} are the data structure used in the API interactions. The {\em Generated Services} provide function calls that are used to perform GET, POST, etc operations on the remote server. All services inherit from \texttt{AbstractService}, which is a class that is added to all generated solutions. 
\texttt{AbstractService} handles network logic, and moves the low-level networking code out of the files that the developer can decide to modify. All the service requests return a \texttt{Response} object, which contains an HTTP status code and a data structure ({\em Generated Model}) of a type that depends on the endpoints return type. 

\begin{figure}[b!]
\begin{verbatim}
[env:esp32]
platform = espressif32
board = nodemcu-32s
framework = arduino
lib_deps =
    github.com/steinwurf/bourne.git
extra_scripts = 
    pre_compiling_bourne.py
\end{verbatim}
\caption{Implementation of ESP32 environment in \texttt{platformio.ini}}
\label{fig:ESP32_environment}
\end{figure}

As illustrated in Figure \ref{fig:architecture} the API client is generated based on an OAS, which can support different media types such as JSON, XML, and Images~\cite{OpenAPIMediaTypes}. JSON is the only media type implemented in \texttt{cpp-tiny-client}. %The API client serializes and deserializes JSON objects by means of the \texttt{Bourne JSON} library.
\texttt{cpp-tiny-client} support the HTTP request methods \texttt{GET}, \texttt{DELETE}, \texttt{PUT} and \texttt{POST}, and the request parameters can be specified as \texttt{query} (e.g.: /users{\bf?role=admin}), \texttt{path} (e.g.: /users/{\bf\{id\}}), \texttt{body} (used for \texttt{POST} methods and more rarely for \texttt{PUT} methods), and \texttt{form} (used sometimes for \texttt{POST} methods, and to send binary files) -- meaning that \texttt{header} parameters are not supported yet.

An TLS connection is necessary for HTTPS. To validate the identity of the server and establish the TLS connection, the IoT device needs the CA root certificate corresponding to the certificate of the server. Because of the flash memory constraint, it is difficult to store many root certificates on the microcontroller, and the user has to select manually one or more certificates to be included in the PlatformIO pipeline. 

When generating an API client with \texttt{cpp-tiny-client}, it is possible to specify the targeted microcontroller (the default microcontroller is ESP32), and the generated code will be tailored for the specified IoT device. 
The file \texttt{platformio.ini} is required by PlatformIO to specify the microcontroller environment and external libraries such as Steinwurf's Bourne library, and it is generated based on the board\footnote{\texttt{nodemcu} for ESP32 and \texttt{d1-mini} for ESP8266} selected for the 
code generation. 
An example of \texttt{platformio.ini} file for the ESP32 is shown in Figure~\ref{fig:ESP32_environment}. The file can be modified by the developer to match different hardware.

The ESP32 and ESP8266 use different HTTP libraries. The generator outputs the proper code for either HTTP or HTTPS and setup the \texttt{platformio.ini} configuration file, based on the targeted microcontroller. HTTPS is supported on ESP32. The ESP8266 uses x509 fingerprints \cite{Esp8266HTTPClient_Example} for TLS connections, and its support is not implemented yet.

\begin{figure}[t!]
    \centering
\begin{forest}
    for tree={font=\sffamily, grow'=0,
    folder indent=.9em, folder icons,
    edge=densely dotted}
    [cpp-tiny-client
      [lib
          [models]
          [services]
          [TestFiles]
      ]
      [src
          [main.cpp, is file]
      ]
      [test
      [RunTests.cpp, is file]
      ]
      [platformio.ini, is file]
      [README.md, is file]
      [root.cert, is file]
      [pre\_compiling\_bourne.py, is file]
    ]
\end{forest}
\caption{Folder structure of API client}
\label{fig:folder_structure}
\end{figure}
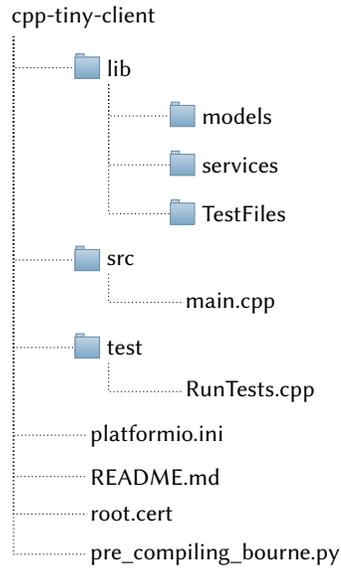

Figure~\ref{fig:folder_structure} reports on the structure of the code for the generated API client. 
The files in the \texttt{models} folder describe non-primitive data types used for the HTTP requests and response. \texttt{services} contains methods to interact with API endpoints on a server. 
The \texttt{root.cert} file is the copy of the certificate fed by the developer. It is possible to ship to the IoT device more than one certificates, whose names will be \texttt{root1.cert}, \texttt{root2.cert}, etc.
The file \texttt{platformio.ini} is used by PlatformIO to specify the microcontroller environment and external libraries during the compilation process. 
The files in the \texttt{TestFiles} folder are used to validate generated code, and as examples for the usage of the generated library. 
Finally, since \texttt{Bourne JSON} uses 11 C++ functions that are not compatible with Espressif, the \texttt{pre\_compiling\_bourne.py} file gets executed by PlatformIO at compile time to "fill the gaps" and to make \texttt{Bourne JSON} compatible with Espressif.

\section{Implementation}\label{sec:implementation}
The core of OpenAPI Generator is composed of an engine and a set of Mustache template files~\cite{Mustache5}, which use placeholder tags that can be replaced with zero, one or multiple values given a context. 

The OpenAPI Generator engine is written in Java. 
The codebase contains a code generator class for each language (python, java, etc) or framework (flask, spring, etc) that OpenAPI Generator targets, plus a few common classes used for example to preprocess the OAS. 

The OpenAPI Generator engine parses OASs as tree data structures, and moves between its nodes while processing the Mustache templates.
Let us consider that a OAS was parsed into the following data structure, represented as JSON for sake of readability:
\begin{verbatim}
{
  "city": "Madrid",
  "museum": [
    { "name": "el Prado" },
    { "name": "Cerralbo" },
    { "name": "Sorolla" }
  ]
  "x-hot": true
}
\end{verbatim}

The following Mustache template:

\begin{verbatim}
The city {{name}} has
the {{#museum}}{{name}} Museum, {{/museum}}
and much more.
{{#x-hot}}
And it's hot in the Summer!
{{/x-hot}}
\end{verbatim}

will produce the following: 

\begin{verbatim}
The city Madrid has
the el Prado Museum,
the Cerralbo Museum,
the Sorolla Museum,
and much more.
And it's hot in the Summer!
\end{verbatim}

%\texttt{cpp-tiny-client} is implemented using Mustache templates, which are logic-less templates \cite{Mustache}. For this project, we have mainly been working with the Mustache constructs: \texttt{Sections}, \texttt{Inverted Sections}, \texttt{Non false values} and \texttt{Non-Empty Lists} as explained in \cite{Mustache}. These constructs cover \texttt{If}, \texttt{Not if}, \texttt{Objects} and \texttt{For loops}. 

\begin{figure}[b!]
\begin{verbatim}
begin(std::string url)
{
    {{#isESP8266}}
    http.begin(client, url);
    {{/isESP8266}}

    {{#isESP32}}
    http.begin(url, root_ca);
    {{/isESP32}}
}
\end{verbatim}
\caption{Implementation of network logic in AbstractService.h}
\label{fig:AbstractService}
\end{figure}

All generators are based on the \texttt{DefaultCodegen} class, which provides hooks for each part of the generation process. 
The users of \texttt{cpp-tiny-client} can specify the targeted microcontroller through a Command Line Interface (CLI) option, currently the available options are  \texttt{ESP32} and \texttt{ESP8266}, and the option is parsed through the \texttt{cliOptions} hook. 
This option is used in \texttt{AbstractService} to differentiate between the network logic for the two microcontrollers. The services use the same method \texttt{begin}, inherited from \texttt{AbstractService}, to initializing a request, but it is populated with a different network logic based on the microcontroller at hand. 
Figure~\ref{fig:AbstractService} shows a simplified code snippet of how Mustache is used to differentiate between the two controllers.  %Generated code for both of them connect to a server given an \texttt{URL}. 
The interface of the ESP8266 needs a network client to make an HTTP handshake, whereas the ESP32 handles network connections internally, and makes the HTTPS handshake with a certificate given as input.

%The variable \texttt{root\_ca} comes from \texttt{AbstractService}. It is included as a preprocessor, from a file called root.cert. %, as shown in Figure \ref{fig:root_cert_include}.

%\begin{figure}[!h]
%\begin{verbatim}
%const char* root_ca =
%#include "../../root.cert"
%;
%\end{verbatim}
%     \caption{Include of root certificate in AbstractService.h}
%     \label{fig:root_cert_include}
% \end{figure}
%%platformio

The CLI options are also used to set the microcontroller environment in \texttt{platformio.ini}. A PlatformIO environment for the ESP32 is shown in Figure \ref{fig:ESP32_environment}. 
The \texttt{platform}, \texttt{board} and \texttt{framework} properties need to be set based on the targeted microcontroller. The \texttt{lib\_deps} property is used to import the \texttt{Bourne JSON} library. 

Espressif controllers use an older version of C++, which clashes with what the \texttt{Bourne JSON} library expects. To make the latter compatible with the compiler, the Bourne library was modified with a \texttt{\#define \_GLIBCXX\_USE\_C99 1} to \texttt{parser.cpp} and \texttt{json.cpp}. This and other operations are performed by the \texttt{pre\_compiling\_bourne.py} script, which visits the two files and modify them for the sake of compilation against the Espressif controllers. 

To add support for a new microcontroller to \texttt{cpp-tiny-client}, (i) the name of the controller needs to be defined in \texttt{cliOptions} and \texttt{processOpt}; (ii) the network logic for the new microcontroller needs to be implemented in \texttt{AbstractService.h} and in the service templates; and (iii) a \texttt{platformio.ini} must be generated to specify \texttt{platform}, \texttt{board} and \texttt{framework} properties.

%The body of the requests and responses is represented as a string following the JSON format. To save the user from implementing logic to transform models to and from JSON when communicating with the server, each generated model implements a \textit{toJson} and \textit{fromJson} method. Both these methods iterate over the model variables, either instantiating them, based on values in a JSON object, or converts them and their values to a JSON object. Both methods uses built-in functionalities in the Bourne JSON Library to differentiate between types, nested elements, etc. 

%If a service method is specified in the OpenAPI Specification to return the map datatype, we change it to a string.

%%%%%%%%%%%%%%%%%%%%%%%%%%%%%%%%%%%%%%%%%%%%%%%%%%%%%%%%%%%%%%%%%%%%%%%%

\section{Validation}\label{sec:validation}

\begin{figure}[b!]
    \centering
    \includegraphics[width=0.6\linewidth]{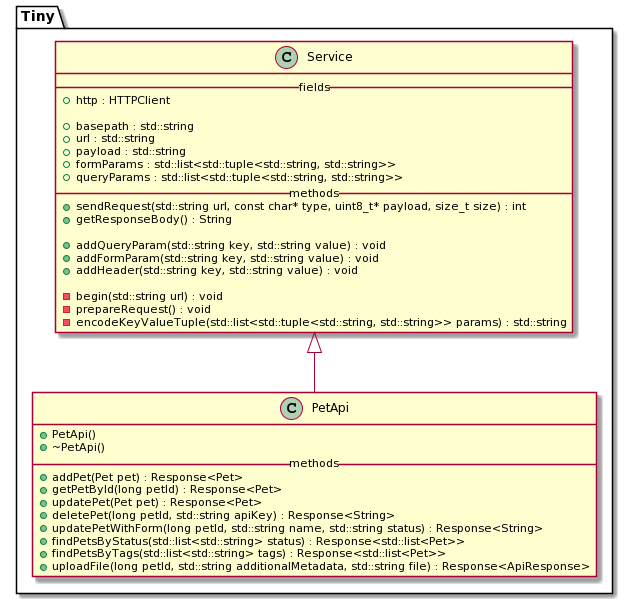}
    \caption{UML Class diagram for the  \texttt{PetApi} service}
    \label{fig:testapi}
\end{figure}

To validate the correctness of the generated API client, we considered both the generated models and services. %For the models, unit tests are generated through templates, to test serialization and deserialization of primitive data types. 
Subsection~\ref{subsec:apitest} details how we used the official OpenAPI's test server to test the generated client code. Subsection~\ref{subsec:footprint} analyzes the memory footprint of the generated code.

\subsection{API Test}\label{subsec:apitest}

\begin{figure}[t!]
    \centering
    \includegraphics[width=6cm]{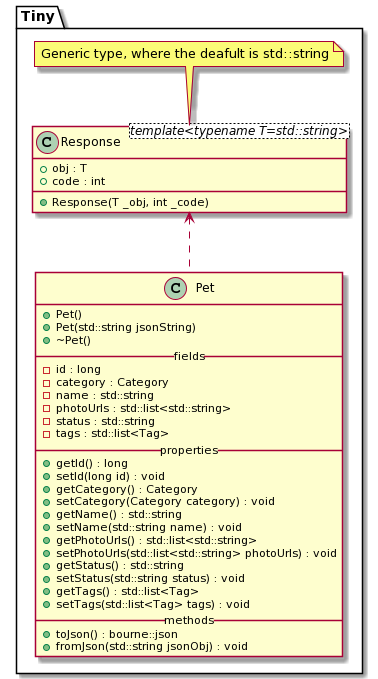}
    \caption{UML Class diagram for the  \texttt{Pet}  model}
    \label{fig:testmodel}
\end{figure}

The validation code was generated based on the \texttt{Petstore} OAS, and the generated code was validated against the official OpenAPI \texttt{Petstore} test server\footnote{https://petstore3.swagger.io/}. 
Figure~\ref{fig:testapi} and Figure~\ref{fig:testmodel} show the UML class diagrams for one of the models and one of the services respectively.

From the \texttt{Petstore} OAS, a \texttt{PetService}, \texttt{OrderService} and \texttt{UserService} classes were generated. 
The tests were performed by calling methods in these generated services, and by comparing the output of generated code with the result of API requests using the CURL tool~\cite{CURL}.

\begin{table}[b!]
\resizebox{0.7\linewidth}{!}{%
\begin{tabular}{|l|l|l|p{0.4\linewidth}|}
\hline
Index & Endpoint             & HTTP Method & Description                                    \\ \hline
1     & /pet                 & PUT         & Update an existing pet                         \\ \hline
2     & /pet                 & POST        & Add a new pet to the store                     \\ \hline
3     & /user/createWithList & POST        & Creates a list of users with given input array \\ \hline
4     & /pet/\{petId\}       & DELETE      & Deletes a pet                                  \\ \hline
5     & /pet/\{petId\}       & GET         & Find pet by ID                                 \\ \hline
\end{tabular}%
}
\caption{Description of tested endpoints}
\label{tab:endpoints}
\end{table}

Table \ref{tab:endpoints} describes the endpoints used for testing, and Table~\ref{tab:expected} describes the input each service method is called with and the expected output from the server response.
\begin{table}[ht!]
\resizebox{0.5\linewidth}{!}{%
\begin{tabular}{|l|l|l|l|}
\hline
Index & Input                                                               & Expected Output                                                           & Parameters \\ \hline
1     & Pet object                                                          & Pet object                                                                & Body              \\ \hline
2     & Pet object                                                          & Pet object                                                                & Body              \\ \hline
3     & List$<$User$>$ & List$<$User$>$       & Body              \\ \hline
4     & int64 petId                                                         & String                                                                    & Path param        \\ \hline
5     & int64 petId                                                         & String                                                                    & Path param        \\ \hline
\end{tabular}%
}
\caption{Input and output for each endpoint}
\label{tab:expected}
\end{table}
The test code was executed on an ESP32, and the server response was logged in the terminal. 
By comparison with the execution of the CURL tool, it was possible to assert that the code was behaving correctly.

\subsection{Memory footprint}\label{subsec:footprint}
As previously discussed, the memory consumption on microcontrollers is of concern. Thus, we analyzed the memory footprint of our API client generated for an ESP32, using the PlatformIO's memory analysis tool. 

\begin{figure}[b!]
    \centering
    \includegraphics[width=0.6\linewidth]{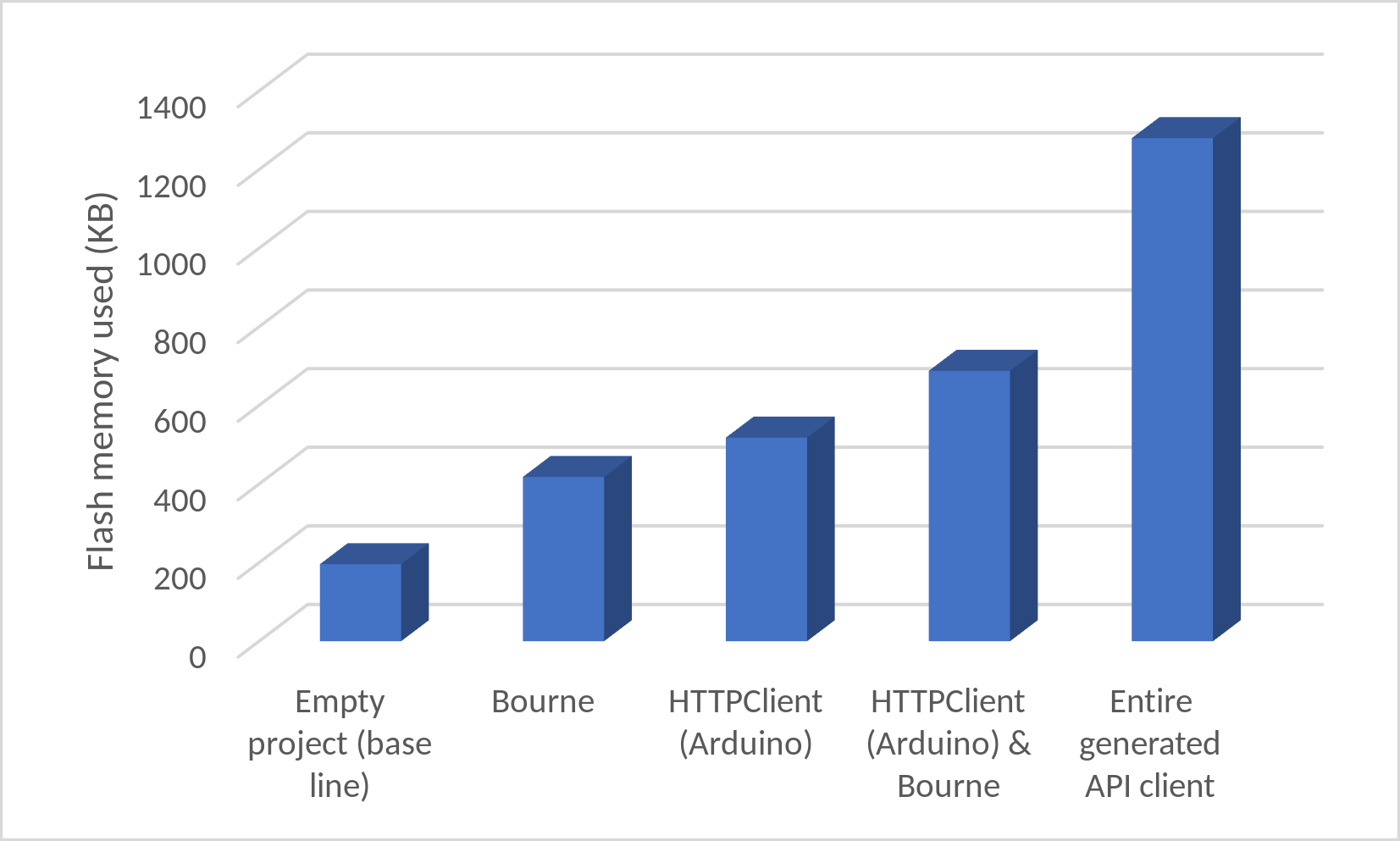}
    \caption{Memory footprint against what got included into the source code}
    \label{fig:footprint}
\end{figure}

Figure~\ref{fig:footprint} shows the flash memory footprint for a simple project using the different libraries used for the ESP32. 
Numerical data are also reported in Table~\ref{tab:memory_used}. 
An empty project occupies 196 KB. 
A project with \texttt{Bourne JSON} and \texttt{HTTPClient} weighs 688 KB, and an API client generated with the petstore OAS uses 1280 KB, hinting that the code generated by \texttt{cpp-tiny-client} has a small footprint, and that 4 MB of flash memory of an ESP32 is enough to include reasonably complex business logic. 
%
%This generated client uses almost twice the flash memory, as a project including only the two libraries. 
However, the tests show that the generated code can only be used by microcontrollers whose memory is at least 688KB. 

%%%%%%%%%%%%%%%%%%%%%%%%%%%%%%%%%%%%%%%%%%%%%%%%%%%%%%%%%%%%%%%%%%%%%%%%

\section{Conclusions \& Future Work}\label{sec:limitation}

This report documents the design and implementation of \texttt{cpp-tiny-client}, an API client generator for IoT devices. The generator is a plugin for the OpenAPI Generator project, and it is compatible with ESP32 and ESP8266 microcontrollers. 
\texttt{cpp-tiny-client} uses PlatformIO as build platform, and the generated code uses Espressifs' \texttt{HTTPClient} and \texttt{Esp8266HTTPClient} libraries for HTTP requests and \texttt{Bourne JSON} for serialization and deserialization purposes.

\begin{table}[t!]
\resizebox{0.5\linewidth}{!}{%
\begin{tabular}{|l|l|}
\hline
Contains                       & Flash memory used (KB) \\ \hline
Empty project (base line)      & 196                    \\ \hline
Bourne                         & 418                    \\ \hline
HTTPClient (Arduino)           & 518                    \\ \hline
HTTPClient (Arduino) \& Bourne & 688                    \\ \hline
Entire generated API client   & 1280                    \\ \hline
\end{tabular}%
}
\caption{Memory consumption overview}
\label{tab:memory_used}
\end{table}

%% Link til github repo, evt til readme hvordan man bruger den. Det så jeg i et andet paper

% \texttt{cpp-tiny-client} was accepted into the main branch of OpenAPI Generator\footnote{\url{https://github.com/OpenAPITools/openapi-generator/pull/9489}}, and is in beta for evaluation by users. 

\texttt{cpp-tiny-client} is in its beta phase for evaluation by users. 
Future work aims to support HTTPS for ESP8266, by distilling a x509 fingerprint from certificates, as needed by the ESP8266HTTPClient library. 
Some more work is required to support request parameters encoded as headers, and use it for basic authentication support. 
%
%% Media types
Currently, the API clients generated by \texttt{cpp-tiny-client} can only use the JSON media type. We plan to add native support for other media types, such as the image media type, to cover for example the use case of a IoT device controlling a camera and posting collected images to a remote server.

%Only supports ESP32 and ESP8266
The current structure of \texttt{cpp-tiny-client} hints that it is possible to support further microcontrollers. 
With the vast differences in microcontroller architectures, available RAM and flash memory, it is not likely that the same API client can be compatible with all platforms. Rather, future work would consist of creating more sub-generators, each supporting their own branch of microcontrollers, for example by including the libraries necessary to support the particular architecture. 
A place to start, would be to make an MbedOS API client generator, which is a different popular open source framework for microcontrollers~\cite{MbedOS}. MbedOS has an HTTP request library~\cite{MbedOSHTTPRequests}, similar to the HTTP request libraries used in Arduino in \texttt{cpp-tiny-client}. 
%% expand wih mbedos
This effort would provide OpenAPI Generator with a solid foundation for generating API clients for diverse microcontrollers, which can hopefully lead to further development on the code generator. 
%%%%%%%%%%%%%%%%%%%%%%%%%%%%%%%%%%%%%%%%%%%%%%%%%%%%%%%%%%%%%%%%%%%%%%%%

%%%%%%%%%%%%%%%%%%%%%%%% SLUTNINGEN PÅ REGULÆR PAPER %%%%%%%%%%%%%%%%%%%%%
%%%%%%%%%%%%%%%%%%%%%%%%%%%%%%%%%%%%%%%%%%%%%%%%%%%%%%%%%%%%%%%%%%%%%%%%%%

\section*{Acknowledgment}

Research funded in part by the European Union through the Horizon 2020 projects DomOS (grant agreement 894240) and FEVER (grant agreement 864537), and by Innovation Fund Denmark project FED (grant agreement 8090-00069B).

The work was supported by the OpenAPI Generator committee, and in particular by its leader William Cheng, which has evaluated the proposed \texttt{cpp-tiny-client} generator in its initial form. Thanks to Steinwurf and Espressif for their open source libraries. 

%The research presented in this paper was supported in part by the DiCyPS (Center for Data-Intensive Cyber-Physical Systems) project of the Innovation Fund Denmark.
%and it is part of the Validation WP of the ERC Advanced Grant LASSO (Learning, Analysis, SynthesiS and Optimization of Cyber-Physical Systems). 

\bibliographystyle{plain}

\bibliography{references}
%%%%%%%%%%%%%%%%%%%%%%%%%%% -- Appendix -- %%%%%%%%%%%%%%%%%%%%%%%%%%%
\end{document}